\documentclass[lettersize,journal]{IEEEtran}
\usepackage{amsmath,amsfonts}
\usepackage{algorithmic}
\usepackage{algorithm}
\usepackage{array}
\usepackage{textcomp}
\usepackage{stfloats}
\usepackage{url}
\usepackage{caption}
\usepackage{subcaption}
\usepackage{xcolor}
\usepackage{verbatim}
\usepackage{graphicx}
\usepackage{cite}
\begin{document}

\title{TokCom-UEP: Semantic Importance-Matched Unequal Error Protection for Resilient Image Transmission}

\author{Kaizheng Zhang, Zuolin Jin, Zhihang Cheng,~\IEEEmembership{Member,~IEEE,} 
        Ming Zeng, Li Qiao,~\IEEEmembership{Senior Member,~IEEE,} 
        and  Zesong Fei,~\IEEEmembership{Senior Member,~IEEE} 
}



\maketitle

\begin{abstract}
Token communication (TokCom), an emerging semantic communication framework powered by Large Multimodal Model (LMM), has become a key paradigm for resilient data transmission in 6G networks. A key limitation of existing TokCom designs lies in the assumption of uniform token importance, which leads to the adoption of equal error protection (EEP). However, compressed one-dimensional (1D) token sequences inherently exhibit heterogeneous semantic importance hierarchies, rendering EEP schemes suboptimal. To address this, this paper proposes TokCom-UEP, a novel semantic importance-matched unequal error protection (UEP) framework designed for resilient image transmission. TokCom-UEP integrates rateless UEP coding with the non-uniform semantic importance of tokens by partitioning source tokens into nested expanding windows, assigning higher selection probabilities to windows containing critical tokens to ensure their prioritized recovery. Simulation results demonstrate that TokCom-UEP outperforms EEP schemes in terms of three core semantic restoration metrics and spectral efficiency under low-overhead conditions. 
\end{abstract}


\section{Introduction}
\IEEEPARstart{W}{ith} the evolution of sixth-generation (6G) networks and the proliferation of vision-oriented applications, achieving efficient and robust image transmission under adverse channel conditions has emerged as a critical challenge~\cite{6G1,6G2}. As a novel paradigm designed to convey the core meaning of information efficiently, semantic communication is regarded as a promising solution to this problem. Nevertheless, most existing semantic communication frameworks are implemented through end-to-end joint source–channel coding (JSCC). Such a design, however, is difficult to seamlessly integrate into digital communication systems, thereby hindering their deployment in practical networks~\cite{DJSCC}.

Concurrently, the rapid proliferation of Large Multimodal Models (LMMs), with their powerful capabilities in semantic understanding and generation, has injected new momentum into this field. Motivated by the deployment challenges of JSCC and the potential of LMMs, researchers have proposed Token Communication (TokCom)~\cite{TokenCom}. As a novel generative semantic paradigm, TokCom demonstrates two primary advantages. First, it leverages LMMs to convert high-dimensional images into discrete semantic token sequences, achieving a high compression rate while maintaining compatibility with digital systems. Second, it exhibits superior robustness; by utilizing auxiliary multimodal information and contextual dependencies, the receiver can effectively infer and reconstruct tokens that are corrupted or lost during transmission, ensuring reliable semantic delivery under adverse channel conditions.

However, existing TokCom frameworks assume equal token importance, thus adopting equal error protection (EEP) strategies that overlook positional variations in semantic importance. However, insights from image tokenization reveal non-uniform information distribution in compressed 1D token sequence. For example, the work in \cite{hkm} showed that specific token positions exhibit strong correlations with high-level semantic attributes such as background and illumination, demonstrating obvious semantic separation. Crucially, this separation implies that different tokens contribute unevenly to the overall image reconstruction; tokens responsible for dominant visual structures are inherently more critical than those representing subtle details. Simultaneously, the variable-length One-D-Piece model \cite{1dpiece} employs a "tail token dropping" paradigm during training, prioritizing critical semantics in head tokens. This positional dependency underscores EEP's inefficiency in redundancy allocation, rendering semantically vital tokens susceptible to erasures. Motivated by these observations, this work explores token interpretability to devise unequal error protection (UEP) mechanisms within TokCom, enhancing resilience against channel erasures.

In communication systems, channel coding emerges as a prevalent and effective approach to realize UEP. Among such techniques, fountain codes offer a flexible framework for UEP implementation. UEP fountain codes have been widely adopted across diverse applications. For instance, \cite{shang2021expanding} employs feedback-based expanding window fountain (EWF) codes to mitigate packet losses in vehicle-to-everything (V2X) networks for scalable video transmission. Similarly, \cite{yang2017unequal} applies EWF codes to 3D audio delivery, achieving a 14\% improvement in subjective quality compared to EEP counterparts. In \cite{wu2020new}, a novel UEP fountain code variant enhances the efficiency of image transmission in satellite communications, ensuring balanced performance across data partitions. Nonetheless, these conventional designs primarily  operate at the bit or symbol level, overlooking the positional semantic interpretability inherent to tokens. To the best of our knowledge, no prior work has integrated UEP fountain coding with TokCom, presenting an unexplored opportunity to leverage semantic importance-matched rateless encoding for token-based transmission.


In this paper, we introduce TokCom-UEP, a semantic importance-matched UEP framework that tailors channel coding to the non-uniform information density of TokCom. By implementing an EWF-based coding strategy, we effectively rectify the spectral inefficiency of standard EEP approaches, ensuring that semantically critical tokens receive prioritized protection against channel erasures. Extensive simulations reveal that TokCom-UEP yields substantial performance gains in PSNR, LPIPS, and CLIP Score, particularly in low-overhead regimes. 
Crucially, our approach demonstrates significantly improved spectral efficiency, proving that integrating semantic awareness in data trasmission is essential for enhancing robustness and effiency in 6G networks

\begin{figure*}[!t]
    \centering
    \includegraphics[width=\textwidth]{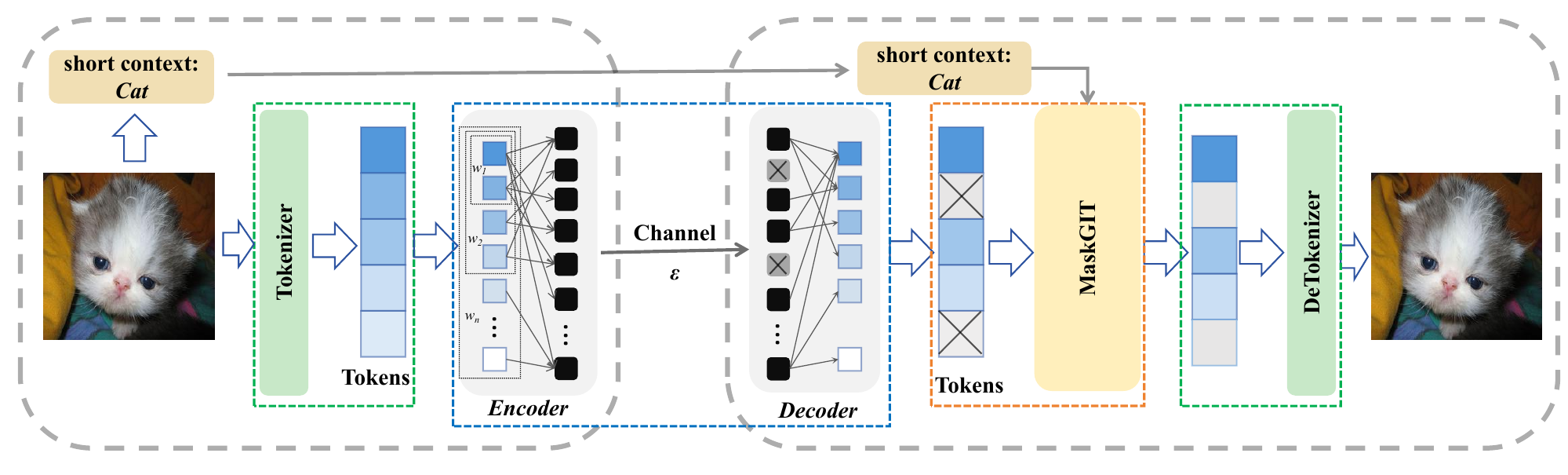}
    \caption{The overall architecture of the proposed TokCom-UEP system. The pipeline proceeds sequentially from Image Tokenization to EWF-based UEP Coding, followed by Multimodal-Guided Token Restoration at the receiver, and finally De-tokenization for image reconstruction.}
    \label{fig:system_model}
\end{figure*}

\section{System Model}
Fig.~\ref{fig:system_model} depicts the proposed TokCom-UEP framework, featuring a modular architecture that integrates semantic tokenization, UEP channel coding, and multimodal reconstruction. The framework comprises three primary modules, with their functionalities elaborated below.

\subsection{Tokenization and DeTokenizer}

As illustrated by the green block in Fig.~\ref{fig:system_model}, the TokCom-UEP framework employs the One-D-Piece model for image tokenization and reconstruction, consisting of a \textbf{Tokenizer} and a \textbf{Detokenizer}. The Tokenizer maps the input image $\mathcal{X} \in \mathbb{R}^{H \times W \times 3}$ to a 1D sequence of discrete tokens $T$ with inherent semantic importance:
\begin{equation}
    T = \text{Tokenizer}(\mathcal{X}) = [T_1, T_2, \ldots, T_L],
\end{equation}
where each $T_i$ denotes a discrete token index drawn from a pre-trained codebook $\mathcal{C}$.
Conversely, the Detokenizer executes the inverse mapping, reconstructing the image $\hat{\mathcal{X}}$ from a prefix subsequence $T' = [T_1, T_2, \ldots, T_k]$ (with $k \leq L$):
\begin{equation}
    \hat{\mathcal{X}} = \text{Detokenizer}(T').
\end{equation}

Notably, during the training phase, the model adopts a ``Tail Token Drop'' strategy, where the Detokenizer is tasked with recovering the original image using only randomly truncated prefixes. This constraint compels the Tokenizer to prioritize encoding the most critical global information into the leading tokens. Consequently, the generated token sequence exhibits a distribution where semantic importance inherently decreases from the head to the tail.

\subsection{EWF-based UEP coding}

\begin{figure}[t]
     \begin{subfigure}[b]{0.45\textwidth}
         \centering
         \includegraphics[width=\textwidth]{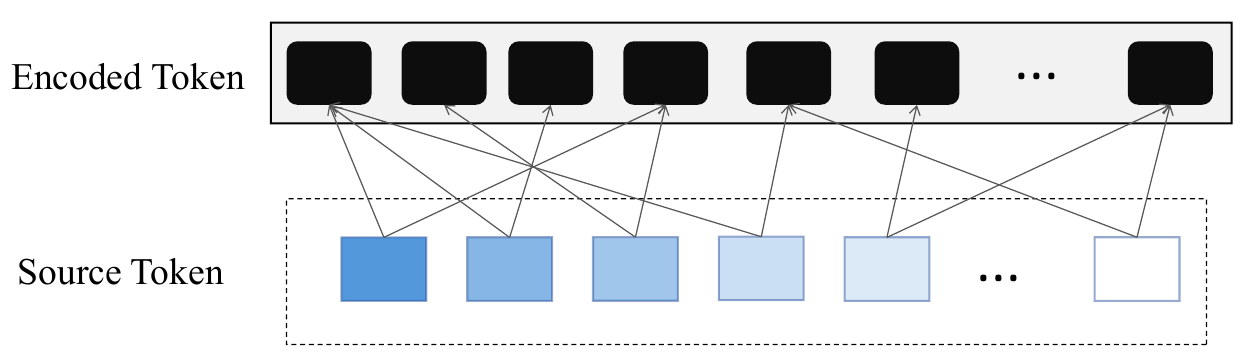}
         \caption{EEP Fountain Code: LT Code.}
         \label{EEP code}
     \end{subfigure}

     \begin{subfigure}[b]{0.45\textwidth}
         \centering
         \includegraphics[width=\textwidth]{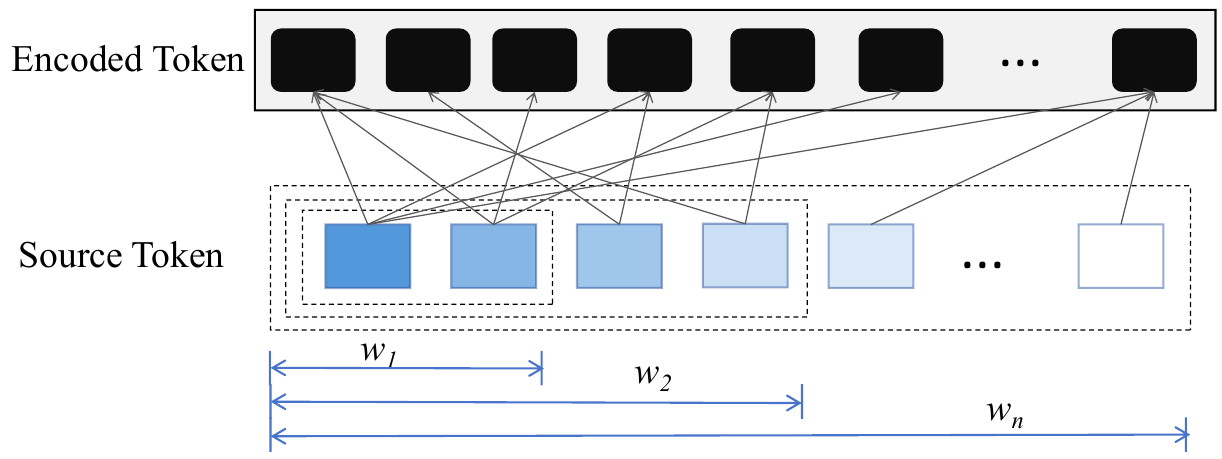}
         \caption{UEP Fountain Code: EWF Code.}
         \label{UEP code}
     \end{subfigure}
        \caption{Example of EEP and UEP Fountain Code.}
        \label{EEPandUEP code}
\end{figure}

As depicted in the blue block of Fig.~\ref{fig:system_model}, the TokCom-UEP framework employs EWF codes to realize UEP within the TokCom paradigm. This rateless coding scheme partitions source tokens into overlapping windows of varying importance, enabling flexible UEP tailored to semantic hierarchies in erasure channels. 
Fig.~\ref{EEPandUEP code} illustrates representative fountain code structures for EEP and UEP: Fig.~\ref{EEP code} depicts an EEP implementation via Luby transform (LT) codes, while Fig.~\ref{UEP code} showcases a UEP variant using EWF codes. Here, each token serves as a source symbol, with encoded tokens generated from source tokens; directed edges denote source token selections during encoding.

Given the sequential decrement in token semantic importance, windows are aligned with token order to enforce UEP commensurate with this hierarchy. As shown in Fig.~\ref{UEP code}, EWF codes utilize nested windows to deliver graduated protection levels, where each window incrementally encompasses prior ones. Consequently, head tokens (with higher semantic salience) exhibit denser connections, implying elevated selection frequencies during encoding and thus superior recovery resilience. In contrast, the LT code in Fig.~\ref{EEP code} imparts uniform priority, yielding equitable selection and recovery probabilities across all tokens.

The EWF-based UEP encoding process unfolds as follows:
\begin{enumerate}
\item \textbf{Window Definition}: Establish nested windows $w_1, w_2, \dots, w_n$, with $w_n = K$ encompassing all $K$ source tokens.
\item \textbf{Window Index Selection}: Sample a window index $j$ from the probability distribution $\Gamma(x) = \sum_{i=1}^r \Gamma_i x^i$, where $\sum_{i=1}^r \Gamma_i = 1$ and $\Gamma_i > 0$; higher $\Gamma_i$ for smaller $i$ prioritizes critical head tokens.
\item \textbf{Degree Sampling}: Sample a degree $d$ from the window-specific distribution $\Omega^{(j)}(x) = \sum_{d=1}^{w_j} \Omega_d^{(j)} x^d$, typically employing the robust soliton distribution (RSD).
\item \textbf{Encoded Token Generation}: Uniformly select $d$ distinct source tokens from the $w_j$ tokens in the chosen window and XOR them to produce an encoded token.
\item \textbf{Iteration}: Repeat steps 2--4 until a sufficient number of encoded tokens is generated.
\end{enumerate}
At the receiver, belief propagation (BP) decoding is applied to recover source tokens from received encoded symbols. Unrecovered tokens are masked with a predefined value (e.g., 4096) for subsequent multimodal inference.

\subsection{Multimodal-Guided Token Restoration}
As illustrated in the orange block of Fig.~\ref{fig:system_model}, at the receiver, a Transformer-based model inspired by the MaskGIT architecture undertakes token restoration post-physical-layer decoding. This module exploits multimodal guidance and contextual inference to enhance resilience against residual erasures. For the decoded token sequence $T_{\text{error}}$ harboring masked positions, the model integrates dual information streams for iterative restoration:
\begin{enumerate}
\item Surrounding uncorrupted tokens, furnishing proximate contextual cues.
\item High-level semantic priors, encapsulated in a concise text prompt $C_{\text{short}}$.
\end{enumerate}

This fusion enables masked token prediction, yielding a semantically coherent corrected sequence $T_{\text{corrected}}$:
\begin{equation}
    T_{\text{corrected}} = \text{MaskGIT}(T_{\text{error}}, C_{\text{short}}).
\end{equation}
Subsequently, $T_{\text{corrected}}$ is input to the Detokenizer for final image reconstruction. This multimodal mechanism augments conventional channel decoding with semantic-aware error correction, providing robust mitigation of information loss in erasure channels.

The three core modules detailed above collectively constitute a holistic semantic importance-matched UEP framework. First, the \textbf{Tokenization and Detokenizer} module establishes a position-dependent semantic hierarchy via the One-D-Piece model, where head tokens encapsulate high-level global semantics and tail tokens represent local details. This intrinsic non-uniformity provides the theoretical premise for the subsequent UEP design. Second, the \textbf{EWF-based UEP Coding} module innovatively aligns nested expanding windows with this token hierarchy, assigning differentiated selection probabilities to ensure prioritized protection for critical semantic tokens. Finally, the \textbf{Multimodal-Guided Token Restoration} module further bolsters system robustness following physical-layer decoding, effectively mitigating residual erasures by fusing contextual cues with high-level semantic priors. These modules operate synergistically: the tokenizer quantifies semantic importance, the channel coder executes differential protection based on this distribution, and the restoration module guarantees high-quality reconstruction even under residual errors. Consequently, this modular design not only preserves the seamless interoperability of TokCom with existing digital communication systems but also significantly enhances transmission resilience in adverse channel conditions through precise semantic importance matching.

\begin{figure*}[t]
     \begin{subfigure}[b]{0.32\textwidth}
         \centering
         \includegraphics[width=\textwidth]{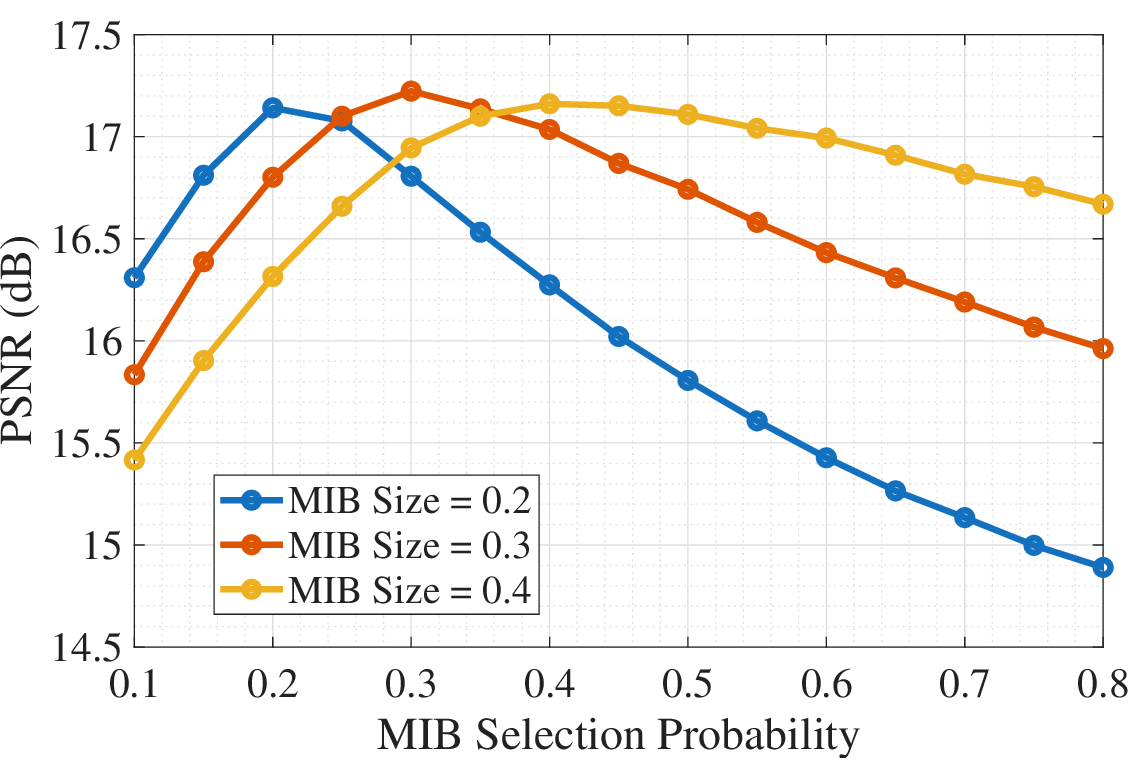}
         \caption{PSNR vs. MIB selection probability}
         \label{PSNRvsMIB}
     \end{subfigure}
     \hfill
     \begin{subfigure}[b]{0.32\textwidth}
         \centering
         \includegraphics[width=\textwidth]{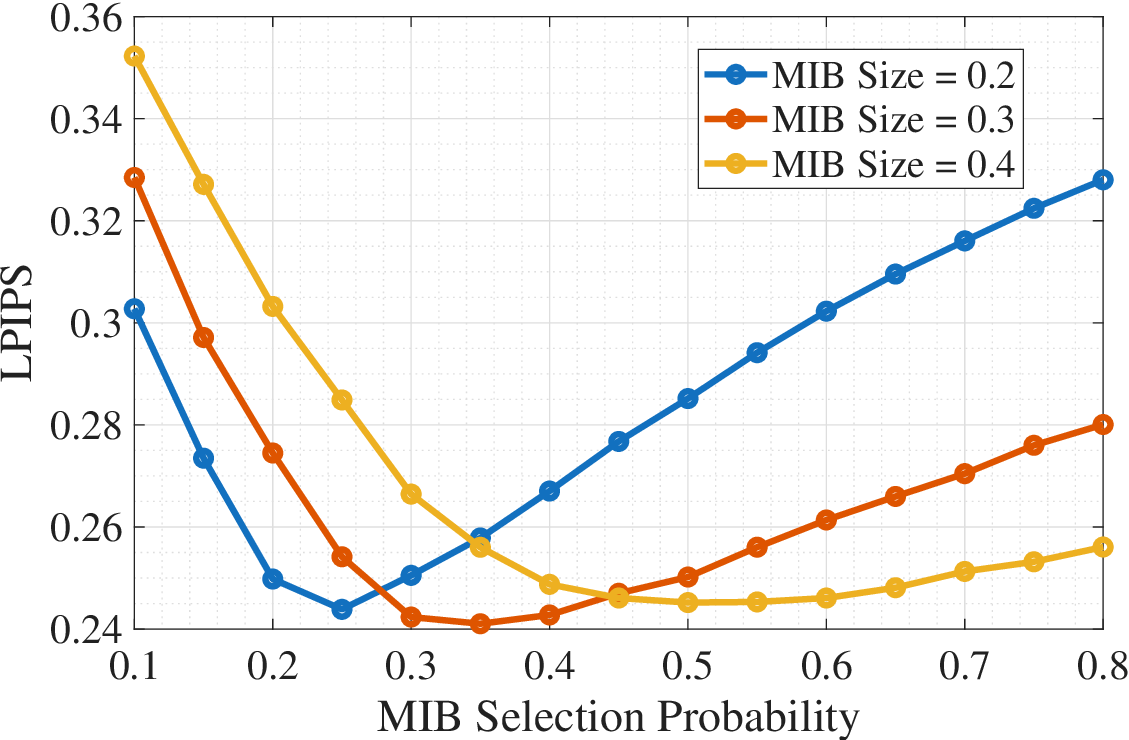}
         \caption{LPIPS vs. MIB selection probability}
         \label{LPIPSvsMIB}
     \end{subfigure}
     \hfill
     \begin{subfigure}[b]{0.32\textwidth}
         \centering
         \includegraphics[width=\textwidth]{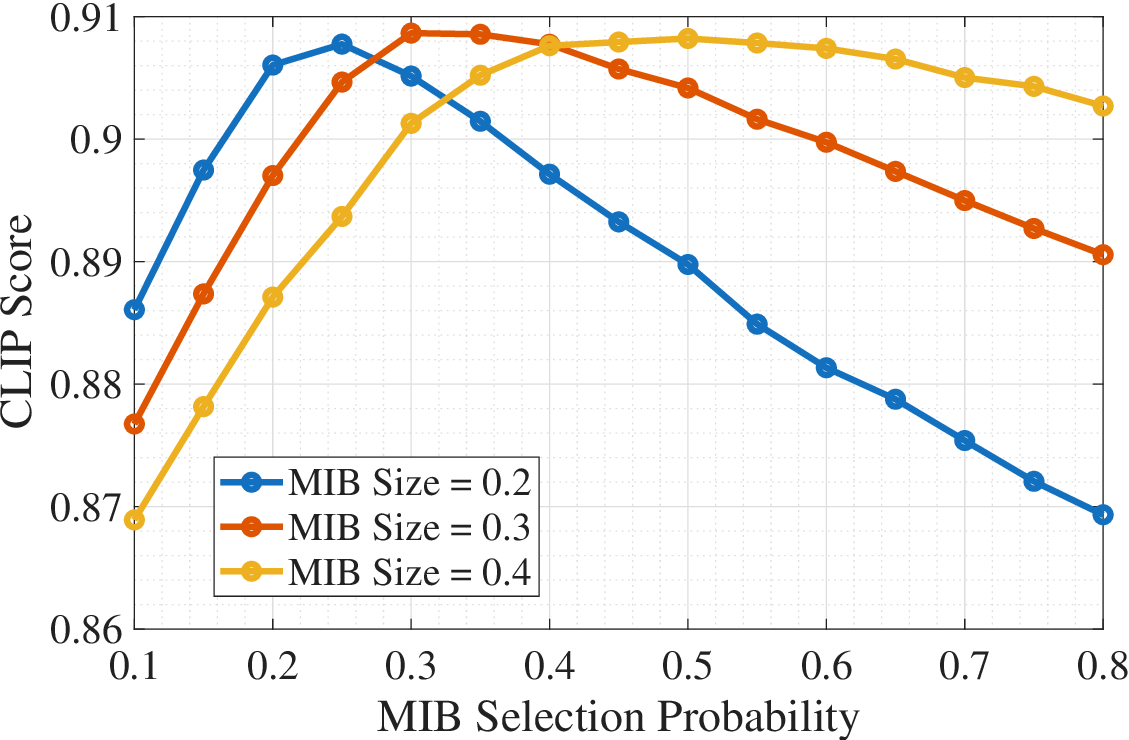}
         \caption{CLIP score vs. MIB selection probability}
         \label{CLIPvsMIB}
     \end{subfigure}
        \caption{Comparison of key performance metrics under different MIB selection probability.}
        \label{performancevsMIB}
\end{figure*}

\begin{figure*}[t]
     \begin{subfigure}[b]{0.32\textwidth}
         \centering
         \includegraphics[width=\textwidth]{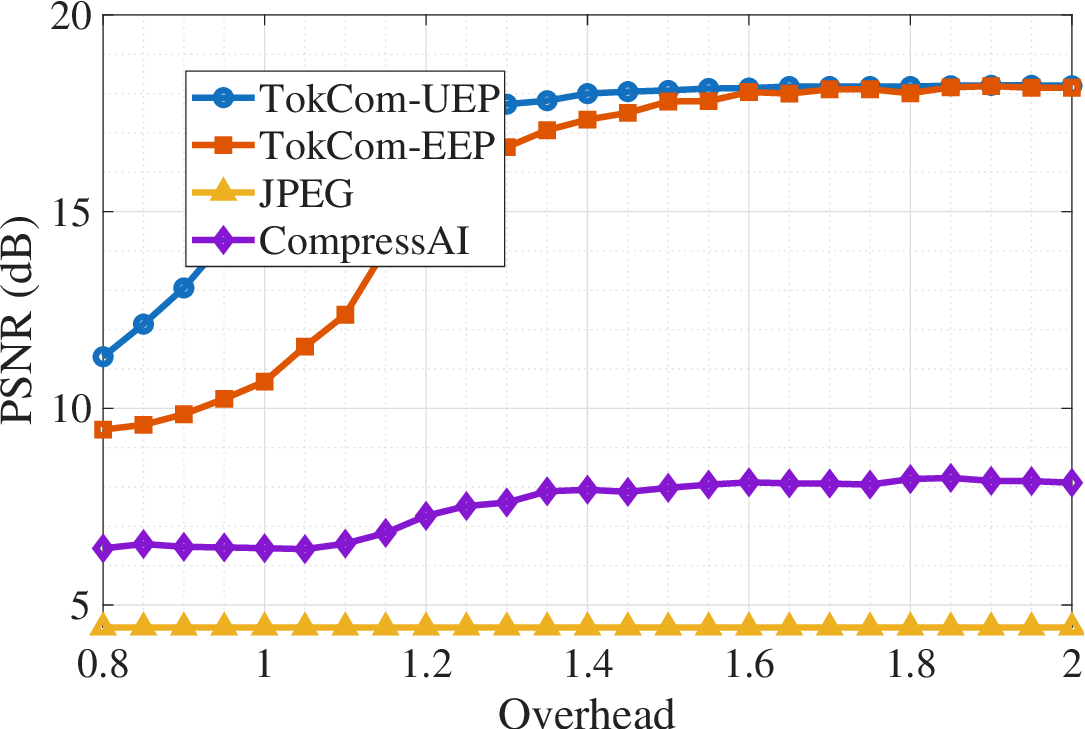}
         \caption{PSNR vs. overhead}
         \label{PSNRvsOverhead}
     \end{subfigure}
     \hfill
     \begin{subfigure}[b]{0.32\textwidth}
         \centering
         \includegraphics[width=\textwidth]{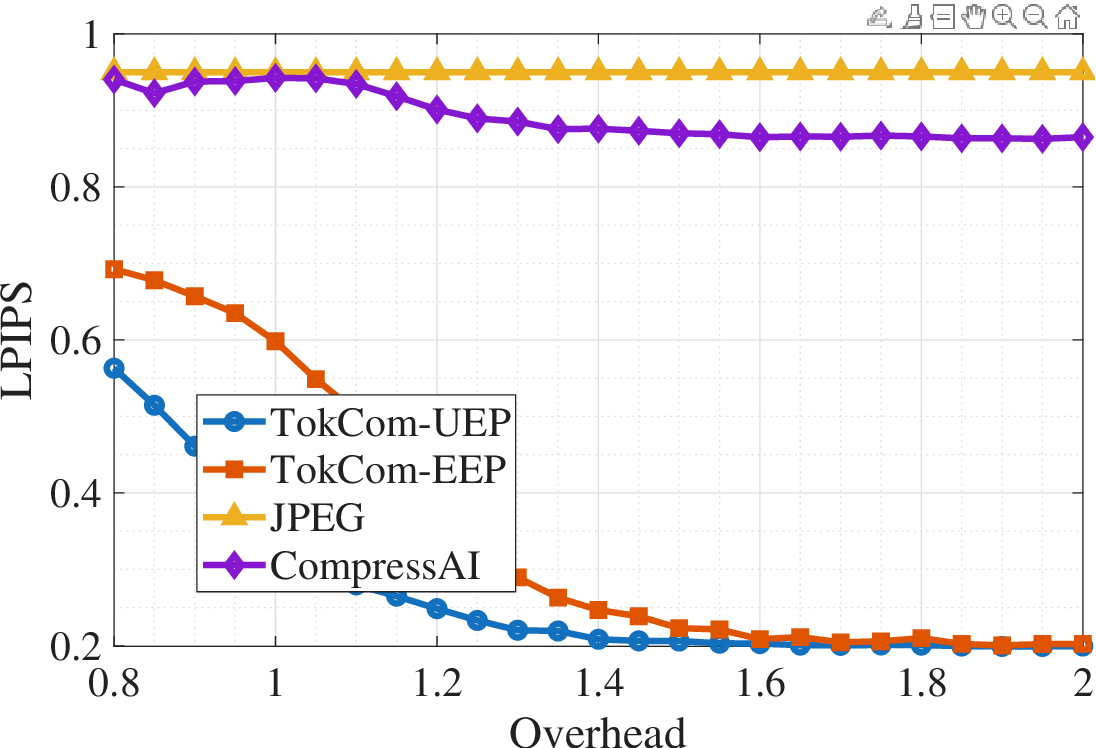}
         \caption{LPIPS vs. overhead}
         \label{LPIPSvsOverhead}
     \end{subfigure}
     \hfill
     \begin{subfigure}[b]{0.32\textwidth}
         \centering
         \includegraphics[width=\textwidth]{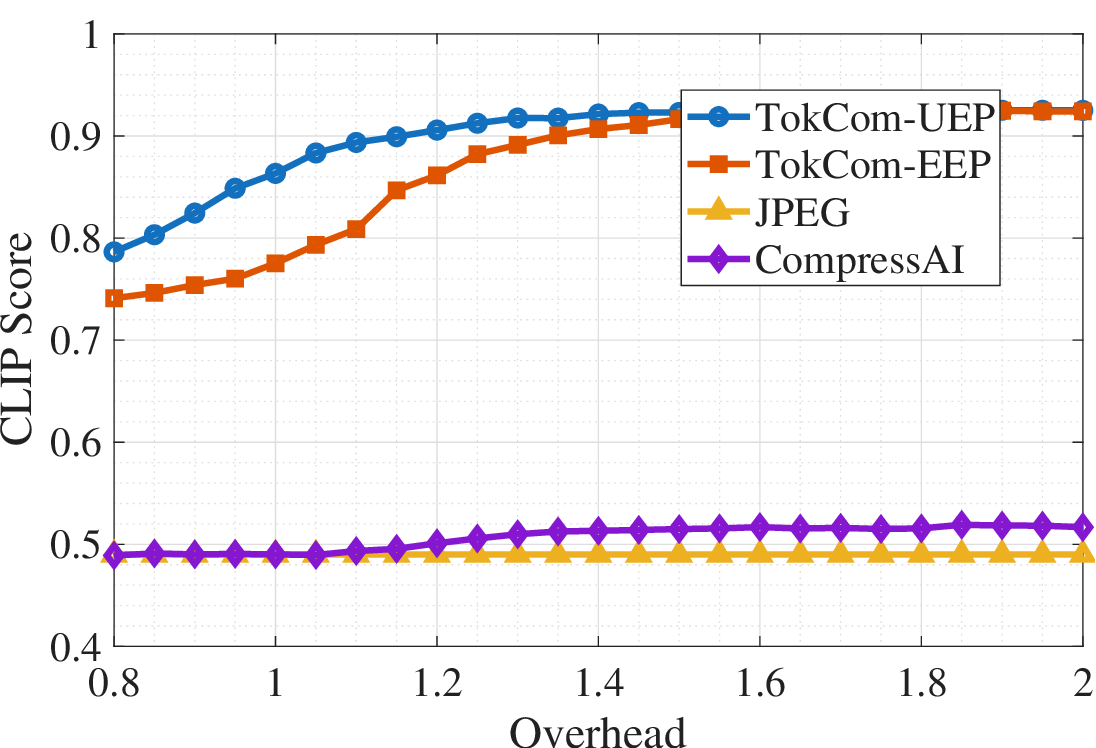}
         \caption{CLIP score vs. overhead}
         \label{CLIPvsOverhead}
     \end{subfigure}
        \caption{Comparison of key performance metrics under different overheads.}
        \label{MetricvsOverhead}
\end{figure*}

\begin{figure}[!t]
    \centering
    \includegraphics[width=0.5\textwidth]{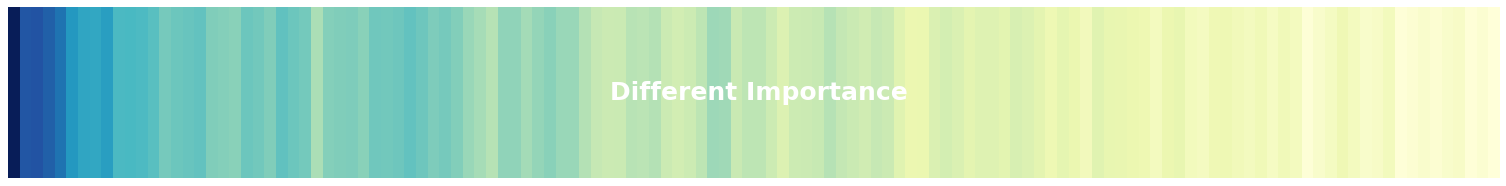} 
    \caption{Importance gradually decreases from left to right. }
    \label{fig:token_importance_distribution}
\end{figure}

\section{Simulation Results}
\subsection{Experimental Setup}
To thoroughly evaluate the performance of the proposed TokCom-UEP scheme, this section details the experimental setup.

\subsubsection{Data and TokCom Model Details}
Experiments are conducted ona carefully selected subset of the ImageNet validation dataset. To mitigate pre-training biases and ensure impartiality, we randomly select 200 distinct classes (with fixed seed=66) and extract 5 images per class, yielding 1000 unique images. Each image is resized to $256 \times 256$ pixels. The semantic communication system leverages a pre-trained One-D-Piece-L-256 model for image tokenization, converting a $256 \times 256 \times 3$ image into a 1D sequence of $L=256$ discrete tokens from a codebook of size 4096. This tokenizer inherently produces a token sequence with position-dependent semantic importance. At the receiver, a Transformer-based MaskGIT model facilitates prediction and restoration of uncorrupted or lost token, guided by the concise semantic prompt $C_{\text{short}}$ for multimodal reconstruction.

\subsubsection{Detailed Communication Setup}
The UEP fountain codes, leveraging an expanding window strategy, are deployed for token transmission over a binary erasure channel (BEC) with erasure probability $p=0.2$. Each token constitutes a source symbol, resulting in $K=256$ symbols partitioned into 2 importance classes. Each coding window adopts the robust soliton distribution (RSD) for degree selection, enabling rateless adaptation to channel impairments.

\subsubsection{Comparative Schemes}
TokCom-UEP is benchmarked against three baselines:
\begin{itemize}
    \item \textbf{TokCom-EEP:} This baseline integrates our TokCom framework with a standard LT code, providing equal error protection for all source tokens using RSD.
    
    \item \textbf{CompressAI:} This scheme combines a state-of-the-art learned image compression model from the CompressAI library, fine-tuned to achieve a bits-per-pixel (BPP) ratio of approximately 0.04, with an LT code for channel protection.
    
    \item \textbf{JPEG:} This traditional baseline utilizes JPEG image compression, adjusted for a BPP of around 0.04, followed by an LT code for channel protection.
\end{itemize}

\subsubsection{Evaluation Metrics}

The performance of all communication schemes is quantified using Peak Signal-to-Noise Ratio (PSNR), LPIPS, CLIP Score and Spectral efficiency. PSNR measures fidelity with higher values indicating better quality. LPIPS correlates with human perceptual judgments, where lower values signify higher perceptual similarity. CLIP Score quantifies semantic fidelity, with higher values indicating better preservation of high-level semantic content. Spectral efficiency is quantified as $\eta = \frac{H \times W \times c}{\text{Encoded Symbols}}$, reflecting the ratio of transmitted encoded symbols to original image pixels, underscoring efficiency in communication.

\subsection{Quantitative Results}

To empirically validate the positional semantic importance in token sequences, we conducted an experiment where we systematically perturbed individual tokens in a sequence by replacing them with random values and then measured the resulting L1 image reconstruction error from the `One-D-Piece` detokenizer. This process was averaged across the validation dataset. As shown in Fig.~\ref{fig:token_importance_distribution}, the results clearly indicate a decreasing trend of semantic importance from head to tail tokens. This non-uniform distribution confirms that initial tokens carry more critical global information, justifying the application of unequal error protection.

Fig. \ref{performancevsMIB} illustrates the variations of three key metrics - PSNR, LPIPS, and CLIP Score - across different values of MIB size $\Pi_i$ and selection probabilities $\Gamma_i$, under a fixed erasure probability of $\epsilon = 0.2$. PSNR and CLIP Score exhibit consistent trends, which stand in inverse relation to that of LPIPS. For each given MIB size, , the peak performance of all metrics consistently occurs at the same selection probabilities. Given the critical role of semantic similarity in the TokCom, this analysis places primary emphasis on the CLIP Score. Notably, we observe that $\Pi_1$ variations modulate both peak efficacy and optimal $\Gamma_1$. Optimal performance emerges at $\Pi_1 = 0.3$ and $\Gamma_1 = 0.3$, where the UEP fountain code enhances erasure resilience and spectral efficiency. These parameters are thus employed in ensuing evaluations.


Fig.~\ref{MetricvsOverhead} depicts PSNR, LPIPS, and CLIP Score metrics for the proposed TokCom-UEP alongside baselines TokCom-EEP, CompressAI-EEP, and JPEG-EEP across varying overheads. Herein, the overhead is defined as the reception overhead, i.e., the fountain code decoding at the receiver is initiated only after receiving $overhead×K$ token symbols. It can be observed that under low overhead scenarios, TokCom-UEP consistently outperforms TokCom-EEP in all three metrics. Specifically, the maximum performance gain of TokCom-UEP over TokCom-EEP is achieved at an overhead of 1: the PSNR gain reaches 40.54\%, the LPIPS gain is 40.15\%, and the CLIP Score gain amounts to 11.36\%. Notably, the performance gap between TokCom-UEP and TokCom-EEP gradually narrows with the increase of overhead, and the two schemes eventually achieve identical performance at an overhead of 1.3. This phenomenon can be attributed to the inherent decoding mechanism of fountain codes: as the overhead increases, the number of recovered tokens by both TokCom-UEP and TokCom-EEP gradually increases, leading to a decreasing proportion of unrecovered tokens relative to the total number of tokens. When the overhead is sufficiently large, both schemes can fully recover all tokens, thus resulting in consistent performance.


\begin{figure}
    \centering
    \includegraphics[width=0.9\linewidth]{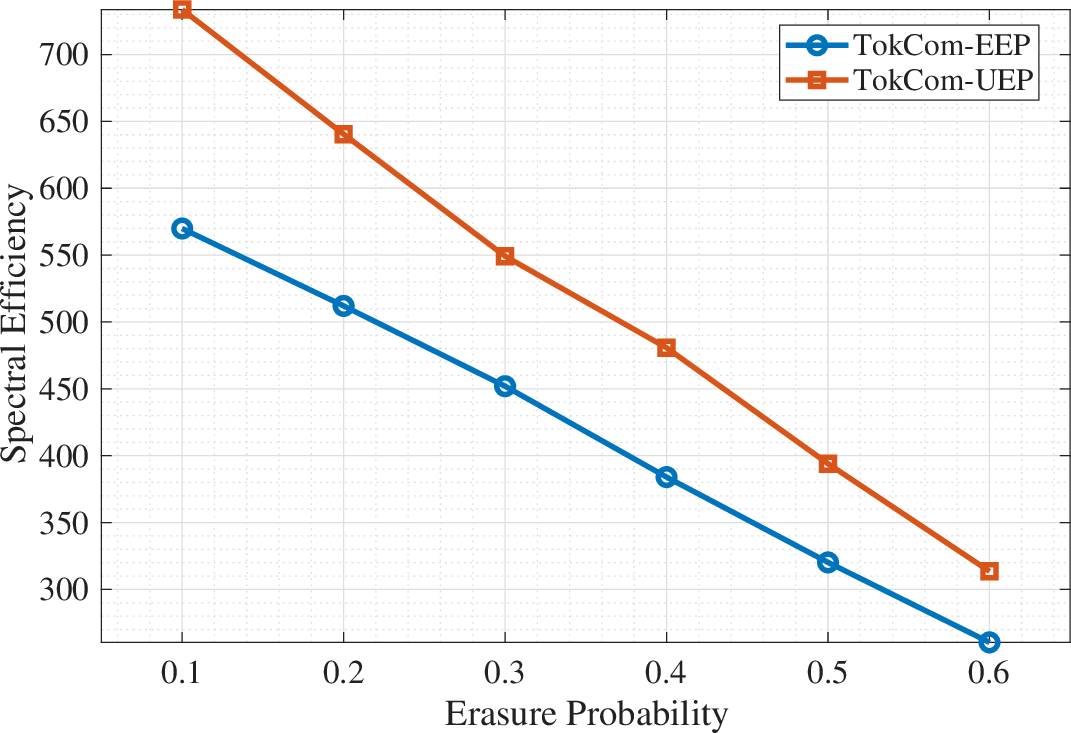}
    \caption{Bandwidth Ratio for CLIP $\ge$ 0.85 under varied erasure probabilities}
    \label{Bandwidth_vs_Erasure}
\end{figure}

To further verify the superiority of the proposed TokCom-UEP scheme in practical communication scenarios, the spectral efficiency performance under different channel erasure probabilities ($\epsilon$) is analyzed, and the results are presented in Fig. \ref{Bandwidth_vs_Erasure}. In this simulation, the reception overhead is fixed at 1.2, and a CLIP Score of 0.85 is defined as the criterion for successful image transmission. Accordingly, the spectral efficiency is calculated as the ratio of the total number of transmitted bits to the total number of pixels of the original image when the CLIP Score reaches 0.85. It can be observed from Fig. \ref{Bandwidth_vs_Erasure} that TokCom-UEP achieves higher spectral efficiency than TokCom-EEP under all tested erasure probabilities. This result fully demonstrates the high efficiency of the proposed TokCom-UEP scheme in adapting to different channel erasure scenarios, which is attributed to its UEP mechanism that prioritizes the protection of semantically important tokens—effectively reducing the redundant bits required to meet the successful transmission criterion.

\section{Conclusion}

This paper investigates the fundamental limitations of existing TokCom frameworks, which overlook the inherent semantic importance hierarchy within 1D token sequences. To address this, we propose TokCom-UEP, a semantic importance-matched UEP scheme specifically designed for token-based image transmission. The proposed TokCom-UEP framework leverages EWF codes to realize semantic importance-matched token transmission, thereby guaranteeing prioritized protection for semantically critical tokens. This approach not only mitigates the suboptimality in EEP-based TokCom variants but also preserves the seamless interoperability of TokCom with conventional digital communication systems. Simulations over the BEC demonstrate the superior performance of TokCom-UEP: at an overhead of 1, it achieves a 40.54\% improvement in PSNR, a 40.15\% enhancement in LPIPS, and an 11.36\% gain in CLIP score compared to TokCom-EEP. Moreover, TokCom-UEP consistently exhibits higher spectral efficiency across varying channel erasure probabilities.

\bibliographystyle{IEEEtran}
\bibliography{main} 

\begin{thebibliography}{1}
\providecommand{\url}[1]{#1}
\csname url@samestyle\endcsname
\providecommand{\newblock}{\relax}
\providecommand{\bibinfo}[2]{#2}
\providecommand{\BIBentrySTDinterwordspacing}{\spaceskip=0pt\relax}
\providecommand{\BIBentryALTinterwordstretchfactor}{4}
\providecommand{\BIBentryALTinterwordspacing}{\spaceskip=\fontdimen2\font plus
\BIBentryALTinterwordstretchfactor\fontdimen3\font minus \fontdimen4\font\relax}
\providecommand{\BIBforeignlanguage}[2]{{%
\expandafter\ifx\csname l@#1\endcsname\relax
\typeout{** WARNING: IEEEtran.bst: No hyphenation pattern has been}%
\typeout{** loaded for the language `#1'. Using the pattern for}%
\typeout{** the default language instead.}%
\else
\language=\csname l@#1\endcsname
\fi
#2}}
\providecommand{\BIBdecl}{\relax}
\BIBdecl

\bibitem{6G1}
Y.~Wang, H.~Han, Y.~Feng, J.~Zheng, and B.~Zhang, ``Semantic communication empowered 6g networks: Techniques, applications, and challenges,'' \emph{IEEE Access}, vol.~13, pp. 28\,293--28\,314, 2025.

\bibitem{6G2}
\BIBentryALTinterwordspacing
M.~S. Akbar, Z.~Hussain, M.~Ikram, Q.~Z. Sheng, and S.~C. Mukhopadhyay, ``On challenges of sixth-generation (6g) wireless networks: A comprehensive survey of requirements, applications, and security issues,'' \emph{Journal of Network and Computer Applications}, vol. 233, p. 104040, 2025. [Online]. Available: \url{https://www.sciencedirect.com/science/article/pii/S1084804524002170}
\BIBentrySTDinterwordspacing

\bibitem{DJSCC}
F.~Wang, X.~Chen, X.~Deng, and S.~Lin, ``Deep joint source-channel coding for wireless image transmission: One-model-to-many,'' \emph{IEEE Wireless Communications Letters}, vol.~14, no.~5, pp. 1501--1505, 2025.

\bibitem{TokenCom}
L.~Qiao, M.~B. Mashhadi, Z.~Gao, R.~Tafazolli, M.~Bennis, and D.~T. Niyato, ``Token communications: A large model-driven framework for cross-modal context-aware semantic communications,'' \emph{IEEE Wireless Communications Magazine}, 2025.

\bibitem{hkm}
\BIBentryALTinterwordspacing
L.~L. Beyer, T.~Li, X.~Chen, S.~Karaman, and K.~He, ``Highly compressed tokenizer can generate without training,'' in \emph{Forty-second International Conference on Machine Learning}, 2025. [Online]. Available: \url{https://openreview.net/forum?id=rcjgzbYvhF}
\BIBentrySTDinterwordspacing

\bibitem{1dpiece}
\BIBentryALTinterwordspacing
Anonymous, ``One-d-piece: Image tokenizer meets quality-controllable compression,'' in \emph{Tokenization Workshop}, 2025. [Online]. Available: \url{https://openreview.net/forum?id=lC4xkcLrdv}
\BIBentrySTDinterwordspacing

\bibitem{shang2021expanding}
X.~Shang and H.~Wu, ``Expanding window fountain code with feedback for svc video transmission in internet of vehicles,'' in \emph{International Conference on Intelligent Transportation Engineering}.\hskip 1em plus 0.5em minus 0.4em\relax Springer, 2021, pp. 190--202.

\bibitem{yang2017unequal}
C.~Yang, R.~Hu, Y.~Song, L.~Su, X.~Wang, and W.~Chen, ``Unequal error protection based on expanding window fountain for object-based 3d audio,'' \emph{Wuhan University Journal of Natural Sciences}, vol.~22, no.~4, pp. 323--328, 2017.

\bibitem{wu2020new}
S.~Wu, Q.~Guan, and Z.~Miao, ``A new class of lt-based uep rateless codes for satellite image data transmission,'' \emph{AEU-International Journal of Electronics and Communications}, vol. 122, p. 153256, 2020.

\end{thebibliography}

\end{document}